# Gate-Variable Mid-Infrared Optical Transitions in a $(Bi_{1-x}Sb_x)_2Te_3$ Topological Insulator

[1]William S. Whitney, [2,3]Victor W. Brar, [4]Yunbo Ou, [5,6]Yinming Shao, [2]Artur R. Davoyan, [5,6]D. N. Basov, [7]Ke He, [7]Qi-Kun Xue and [2]Harry A. Atwater

[1]Department of Physics, California Institute of Technology, Pasadena, California 91125, USA
[2]Thomas J. Watson Laboratory of Applied Physics, California Institute of Technology, Pasadena, California 91125, USA
[3]Kavli Nanoscience Institute, California Institute of Technology, Pasadena, California 91125, USA
[4]Beijing National Laboratory for Condensed Matter Physics, Institute of Physics, The Chinese Academy of Sciences, Beijing 100190, China
[5]Department of Physics, University of California-San Diego, La Jolla, California 92093, USA
[6]Department of Physics, Columbia University, New York, NY
[7]State Key Laboratory of Low-Dimensional Quantum Physics, Department of Physics, Tsinghua University, Beijing 100084, China

*Corresponding author: Harry A. Atwater (haa@caltech.edu)

**Abstract:**
We report mid-infrared spectroscopy measurements of an electrostatically gated topological insulator, in which we observe several percent modulation of transmittance and reflectance of $(Bi_{1-x}Sb_x)_2Te_3$ films as gating shifts the Fermi level. Infrared transmittance measurements of gated $(Bi_{1-x}Sb_x)_2Te_3$ films were enabled by use of an epitaxial lift-off method for large-area transfer of topological insulator films from infrared-absorbing $SrTiO_3$ growth substrates to thermal oxidized silicon substrates. We combine these optical experiments with transport measurements and angle-resolved photoemission spectroscopy to identify the observed spectral modulation as a gate-driven transfer of spectral weight between both bulk and topological surface channels and interband and intraband channels. We develop a model for the complex permittivity of gated $(Bi_{1-x}Sb_x)_2Te_3$, and find a good match to our experimental data. These results open the path for layered topological insulator materials as a new candidate for tunable infrared optics and highlight the possibility of switching topological optoelectronic phenomena between bulk and spin-polarized surface regimes.

Topological insulators – narrow band-gap semiconductors that exhibit both an insulating bulk and metallic Dirac surface states – have been found experimentally in the past several years to display a remarkable range of new electronic phenomena.[1-4] In addition, these Dirac surface states have been predicted to host unique and technologically compelling optical and optoelectronic behavior. Some of these effects have been experimentally demonstrated – giant magneto-optical effects, helicity-dependent photocurrents and more – but many others, including gapless infrared photodetection, gate-tunable, long-lived Dirac plasmons and hybrid spin-plasmon modes, remain elusive.[5-16]

One of the most fascinating features of topological insulator systems is the coexistence and interplay of massless Dirac electrons and massive bulk carriers. While systems like the bismuth telluride family of materials are strong topological insulators, they are also structurally two-dimensional, layered Van der Waals semiconductors.[17,18] For technologies like tunable optics, for which the graphene Dirac system is promising, excitations of both Dirac electrons and these low effective mass bulk carriers are equally compelling.[19-22] The low density of states of both of these classes of carriers and availability of thin, gate-tunable films by Van der Waals epitaxy indicate the possibility of highly tunable infrared absorption.[23,24] Furthermore, by tuning the Fermi level of these materials it may be possible to switch dynamically between optoelectronic regimes dominated by spin-polarized topological surface physics and by bulk semiconductor physics.

In this paper, we report a measurement of the infrared reflectance and transmittance of $(Bi_{1-x}Sb_x)_2Te_3$ topological insulator (TI) films while applying a gate voltage to modulate the Fermi level. To allow gated transmittance measurements, we developed an epitaxial lift-off method for large-area transfer of TI films from the infrared-absorbing $SrTiO_3$ growth substrates to thermal oxidized silicon substrates.[25,26] We combine these optical experiments with gated transport measurements and angle-resolved photoemission spectroscopy to identify the mechanism of the observed spectral modulation. This behavior consists of a gate-driven transfer of spectral weight between both bulk and topological surface channels and interband and intraband channels. We propose that the physical bases for these phenomena are Pauli-blocking of bulk interband transitions for higher photon energies and modulation of intraband transitions with carrier density for lower photon energies. We develop a model for the complex permittivity of gated $(Bi_{1-x}Sb_x)_2Te_3$, and find a good match to our experimental data.

**Results:**

**Device Structure.** Our experimental optical setup and device structure are shown in Fig. 1. The topological insulator film sits atop a thermally oxidized silicon substrate, allowing control of its Fermi level by applying a voltage between electrodes on the film and the doped silicon to accumulate or deplete carriers by the field effect.[27] As depicted in Fig. 1, the topological surface state is thought to occupy a 1 – 2 nm region at the top and bottom interfaces of the $(Bi_{1-x}Sb_x)_2Te_3$ film.[28] An optical microscope image and AFM cross-cut are shown for a $(Bi_{1-x}Sb_x)_2Te_3$ film transferred from its growth substrate to thermally oxidized silicon and patterned into an electrically isolated device.

**Gated Mid-Infrared Spectroscopy.** The primary result of this work is the observation of gate-control of inter and intra-band optical transitions in transmittance and reflectance (Fig. 2a,b). Infrared transmittance and reflectance are probed using an infrared microscope coupled to an FTIR spectrometer, while the gate bias is varied. Modulation of transmittance and reflectance of several percent is observed, with respect to their values at zero-bias applied to the silicon gate. In transmittance, two major features are seen. At lower photon energies, transmittance is increased as the Fermi level is increased. At higher photon energies, transmittance is decreased as the Fermi level is increased. Between these features – labelled A and B, respectively, in Fig. 2b – is an isosbestic point that sees no modulation, suggesting a cross-over between two competing effects.

**Transport and Angle-Resolved Photoemission Spectroscopy.** To locate the Fermi level in our films, we measured the sheet-resistance as functions of gate voltage and temperature (Fig. 3a-c). With negative gate bias, sheet resistance is seen to increase as p-type carriers are depleted by the field effect and the Fermi level of our film is increased. Likewise, with positive bias, sheet resistance is seen to decrease as p-type carriers are accumulated and the Fermi level is decreased. In measurements of sheet resistance versus temperature, a transition from metallic to insulating character is seen as the gate bias passes -40 V – indicating that the Fermi level has crossed the bulk valence band edge.[29] Angle-resolved photoemission spectroscopy (Fig. 3c) is used to map the band structure of $(Bi_{1-x}Sb_x)_2Te_3$ in this region. The total carrier density in the film at zero bias, $n_{2D} = 2.5 \cdot 10^{13}$ cm$^{-2}$, is obtained from Hall measurements.

**Discussion:**

From transport measurements, we conclude that unbiased $(Bi_{1-x}Sb_x)_2Te_3$ films are hole-doped, with a Fermi level position slightly below the bulk valence band edge. The modulation of sheet resistance seen with gating indicates that the Fermi level of the entire film is modified by the gate, though some band bending is expected, as is discussed in Fig 4.[8,23] It further suggests that electrostatic gating is forcing the film between regimes where topological surface carriers and bulk carriers are expected to dominate the conductivity, respectively. At 4.2 K, gate-biasing allows the Fermi level to be pushed from below the bulk valence band edge to near the Dirac point. The $R_{sh}$ on-off ratio is much lower than that seen in films of other layered materials with a similar thickness and band gap, such as black phosphorus, consistent with the presence of a conductive topological surface state that shorts the insulating 'off' state of the field effect device.[30]

Given the identified Fermi level position of the film, we suggest that feature A in the optical response at higher photon energies is driven by gate-modulation of interband transitions via population of bulk valence band states with holes. As shown in Fig. 2c, a doped semiconductor has a characteristic effective bandgap defined – for hole-doped samples – by the distance from the Fermi level to the conduction band. In the $(Bi_{1-x}Sb_x)_2Te_3$ system investigated here, this Fermi level shifts with $V_g$, altering the allowed and forbidden optical transitions and hence its band edge optical constants. Similar behavior is seen for electrostatic doping in graphene, and for chemical doping in narrow-band-gap semiconductor materials, in which it is known as the Burstein-Moss effect.[24,31-33] This behavior is seen only in thin films of materials with a low density of states, and indicates possible technological applications for narrow-gap TI materials as optoelectronic modulators. The observed modulation persists at room temperature, albeit with a lower strength. At a temperature of 10 K, as discussed in the Supplementary Information, the band-edge modulation is stronger and sharper – indicative of a narrower Fermi distribution – and an additional feature appears. The resulting change in the optical band gap can be approximated as follows, where $H(E_{BVB} - E_F, T)$ is a Heaviside step-function with a temperature-dependent broadening that accounts for the width of the Fermi-Dirac distribution.[34]

$$\Delta E_G = 2\Delta(E_{BVB} - E_F)H(E_{BVB} - E_F, T) \qquad (1)$$

We propose that feature B in the optical response at lower photon energies is characterized by a change in the intraband absorption associated with both topological surface states and bulk states. While depressing the Fermi level into the bulk valence band will decrease the band-edge interband transition rate – increasing transmittance near the band edge – it simultaneously transfers spectral weight to intraband channels, increasing absorption and decreasing transmittance at lower energies. We suggest that the physical mechanism of this change in intraband absorption is the modulation of carrier density in the film – by as much as 27 percent – via electrostatics. This behavior indicates the possibility of extending the tunable, mid-infrared Dirac plasmons seen in graphene to spin-polarized topological insulator materials.[10,35,36] This conjecture is supported by our transport data, which indicates that the Fermi level is shifting back and forth across the bulk valence band edge, but transmittance can also be modeled directly using measured values and one free fitting parameter. A simple picture of the modulated bulk interband absorption is provided by experimental measurements of the band edge dielectric function, which was determined from infrared ellipsometry measurements of an as-grown $(Bi_{1-x}Sb_x)_2Te_3$ film on sapphire. The change in the band edge dielectric function energy as a function of gate voltage, is modeled by shifting the zero-bias dielectric function by an energy $\Delta E_S$, proportional to the corresponding voltage, such that the dielectric function as a function of gate voltage can be described by a single free parameter.

To model the optical response at small negative Fermi level positions, the topological surface state and bulk carrier densities are first parameterized as a function of gate voltage. From our fit of the absorption-edge energy shifts, a gate voltage of $V_g = +/-45V$ corresponds to a shift of the Fermi level of approximately 28 meV. The observed metal-insulator transition occurs at $V_g \approx 40V$, so the Fermi level at zero bias must be at approximately (28 meV · 40 V / 45V) = 25 meV below the bulk valence band edge. The bulk valence band is observed to be 150 meV below the Dirac point in angle-resolved photoemission measurements, indicating a Fermi level of $E_F = -175$ meV relative to the Dirac point.[17] The topological surface state carrier density can be calculated from this Fermi level by assuming the electronic structure is characterized by the well-known topological surface state dispersion relation.[17,37] We find that $n_{TSS} = k_F^2/4\pi = 4 \cdot 10^{12}$ cm$^{-2}$ for each surface, where $k_F = E_F/\hbar v_F$. Including both surfaces, our topological surface state density / bulk carrier density ratio is found to be $n_{2D,TSS}/n_{2D} = 30\%$. As our films are deeply subwavelength in thickness, we approximate the $(Bi_{1-x}Sb_x)_2Te_3$ film as having a single effective dielectric function that includes contributions from both of these types of carriers, as well as interband absorption, as discussed above. The intraband dielectric functions for the topological surface state and bulk free carriers were treated using Kubo and Drude models, respectively.[38,39]

$$\varepsilon(\omega) = \varepsilon_{\text{interband}}(\omega, E_F, T) + \varepsilon_{\text{intraband, TSS}}(\omega, n_{2D,\text{TSS}}) + \varepsilon_{\text{intraband, bulk}}(\omega, n_{2D,\text{bulk}})$$

$$= \varepsilon_{\text{interband}}(\omega, E_F, T) - \frac{e^2 v_F}{d\hbar\omega\left(\omega + \frac{i}{\tau}\right)}\left(\frac{n_{2D,\text{TSS}}}{2\pi}\right)^{1/2} - \frac{e^2 n_{2D,\text{bulk}}}{dm\omega\left(\omega + \frac{i}{\tau}\right)} \quad (2)$$

This dielectric function model is combined with a simple capacitor model that defines the change in carrier concentration with gate voltage – up to 27 percent at 90 V. The charge on each plate is given by $Q = V_g C$, where the capacitance C is calculated to be 12 nF/cm$^2$ for the 285 nm SiO$_2$ using a standard parallel plate geometry.[40] Combining these elements, we use the transfer matrix method to calculate transmittance (Fig. 2d) through the $(Bi_{1-x}Sb_x)_2Te_3$ film and substrate stack.[41] The modeled values for ΔT/T – based on experimental parameters and a single fitting parameter – yield a close match to our experimental results.

We note that band bending, as described in Fig. 4, adds an additional degree of complexity to this system.[8,23,42] While our transport data indicate that gating modifies the Fermi level of the entire film, the persistence of feature A in 10 K measurements suggests that gating is less efficient further from the silicon oxide interface. We further note two smaller features seen in FTIR spectra. In Fig. 2a,b a small dip in transmittance and reflectance modulation is seen near 8 microns, which we attribute to absorption in the silicon oxide due to the presence of a phonon line. In Fig. 2b, a small peak in transmittance modulation is seen near 3.8 microns. We speculate that this peak may be due to a defect state or subband, and add that it persists in room temperature measurements and is thus unlikely to be excitonic in nature.[12,43-45]

In conclusion, we have experimentally investigated the mid-infrared optical response of $(Bi_{1-x}Sb_x)_2Te_3$ films as the Fermi level position is varied by electrostatic gating. This response is characterized by a gate-driven transfer of spectral weight between both bulk and topological surface channels and interband and intraband channels. We associate the higher photon energy behavior with Pauli-blocking of bulk interband optical transitions, and the lower energy behavior with topological surface and bulk intraband transition rates that vary with their respective carrier concentrations. These results present layered topological insulator materials as a new candidate material for tunable infrared photonics and illustrate the possibilities for switching topological optoelectronic phenomena – tunable, mid-infrared Dirac plasmons, hybrid spin-plasmons and more – between bulk and spin-polarized surface regimes.

**Methods:**

**Sample Preparation.** The 20 nm $(Bi_{1-x}Sb_x)_2Te_3$ films are grown by molecular beam epitaxy on heat-treated 500 μm-thick $SrTiO_3$ (111) substrates, as previously reported.[17] A mixing ratio of x=0.94 is used for this work. Epitaxial lift-off was used to transfer these films to thermal oxidized silicon substrates, as described in the Supplementary Information. Electron-beam lithography (EBPG 5000+) and reactive ion etching ($SF_6$) are used to pattern the film into electrically isolated squares, and Cr/Au contacts (5 nm / 150 nm) are deposited via thermal evaporation to allow gating and Van der Pauw transport measurements.

**Infrared Spectroscopy.** Infrared spectroscopy measurements are performed with a Nicolet iS50 FTIR coupled to a Continuum microscope with a 50 μm spot size. Samples are wire-bonded and mounted in a Linkam vacuum stage for temperature control at 78 K and 300 K. The band-edge optical constants of the $(Bi_{1-x}Sb_x)_2Te_3$ are extracted with a J.A. Woollam IR-VASE infrared ellipsometry system.

**Transport.** Sheet resistance is measured using the Van der Pauw method and a Janis ST-400 liquid helium cryostat for temperature control from 4.2 to 300 K. Carrier densities are measured in an MMR Technologies Hall system.

**Acknowledgements:**
The authors gratefully acknowledge support from the Department of Energy, Office of Science under Grant DE-FG02-07ER46405 and for facilities of the DOE "Light-Material Interactions in Energy Conversion" Energy Frontier Research Center (DE-SC0001293). W.S.W. also acknowledges support from an NDSEG Graduate Research Fellowship. A.R.D acknowledges fellowship support from the Resnick Institute and the Kavli Nanoscience Institute at Caltech. The authors are grateful to Prof. George Rossman for helpful discussions and use of his FTIR facilities.

**Author Contributions:**

W.S.W., V.W.B and H.A.A. conceived the ideas. Y.O. grew the films and W.S.W. fabricated the devices. W.S.W, Y.S. and Y.O. performed measurements. W.S.W and A.R.D calculated the optical model. All authors contributed to writing the paper. D.N.B, K.H., Q.K.X., and H.A.A. supervised the project.

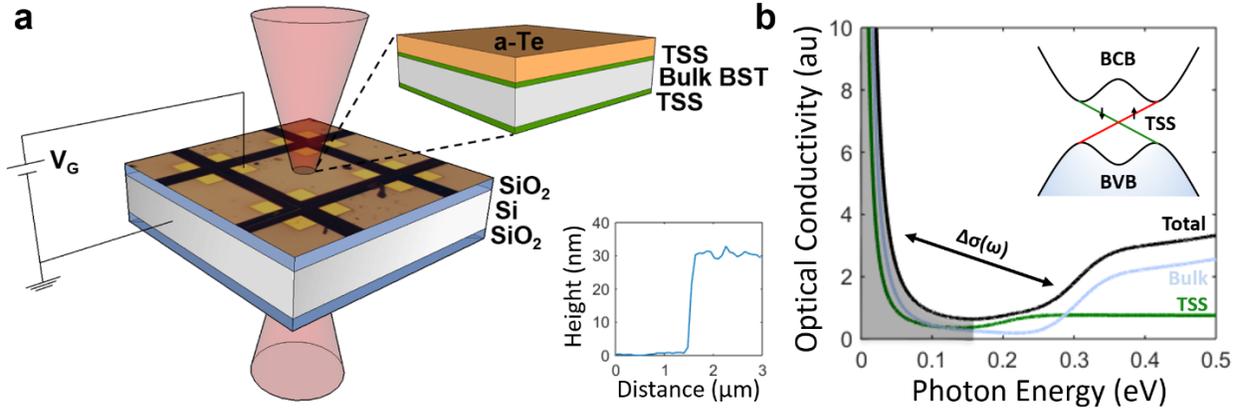

**Figure 1**: Schematic of experiment. **(a)** Schematic of setup. The sample structure consists of a 10 nm a-Te layer atop a 20 nm $(Bi_{1-x}Sb_x)_2Te_3$ (BST) film on 285 nm thermal oxide on silicon. The metallic topological surface states in the BST penetrate 1-2 nm into the insulating bulk. The transmittance and reflectance of this stack are probed by an FTIR spectrometer coupled to an infrared microscope as the gate voltage is modulated. Inset: AFM cross-cut of transferred film, showing 30 nm total height of a-Te and BST. **(b)** Schematic of observed behavior. The optical response of the BST consists of contributions from bulk and topological surface carriers.[46] With changing gate voltage, spectral weight is transferred between both bulk (blue) and topological surface (green) channels and between interband (unshaded region) and intraband (shaded region) channels, as indicated by the arrow. Inset: Schematic band structure of BST, with bulk valence and conduction bands, topological surface states and 0.3 eV band gap indicated.

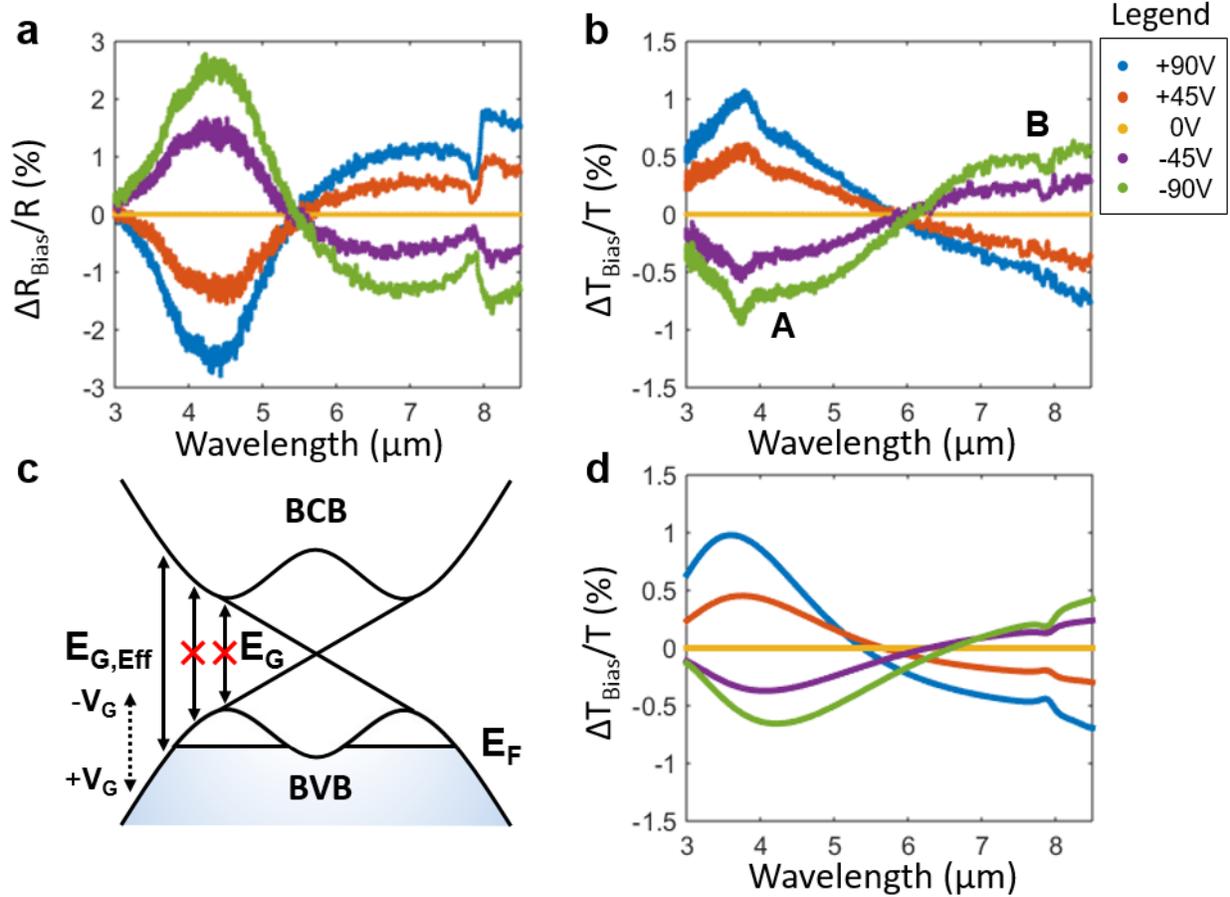

**Figure 2**: Gate-variable FTIR reflectance and transmittance. **(a)** Change in reflectance with electrostatic gate bias at T = 78 K, normalized to the zero-bias case. **(b)** Change in transmittance with gate bias at T = 78 K, normalized to the zero-bias case. Similar behavior is observed at room T, but with lower modulation strength. **(c)** Schematic of the Burstein-Moss effect. As $E_F$ decreases into the BVB, lower energy bulk interband transitions are forbidden. The bulk band gap energy is approx. 300 meV / 4.1 μm. **(d)** Modelled transmittance based on a combined model of gate-variable Pauli-blocking / Burstein-Moss shifting of bulk interband transitions at higher energies and modulation of topological surface and bulk spectral weights at lower energies. As a simple model of the Burstein-Moss shift, band edge optical constants are shifted in energy-space. Varying surface and bulk free carrier contributions to the dielectric function are modelled by the Kubo and Drude models, respectively, and a simple capacitor model of carrier density modulation. From Hall, transport and ARPES results, the zero-bias carrier density is calculated to be 30% topological surface carriers and 70% bulk carriers.

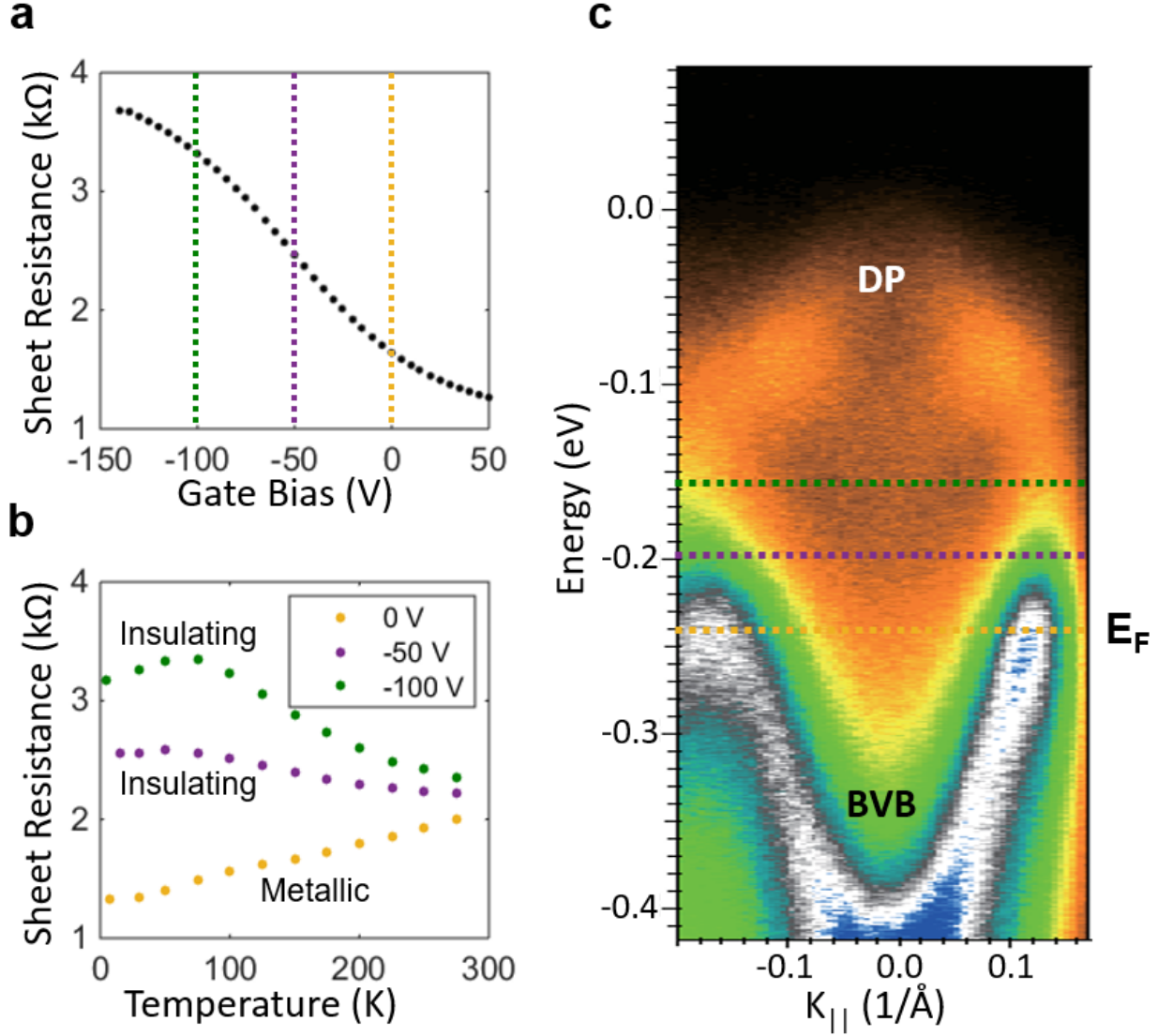

**Figure 3**: Electrical characterization and gate-driven metal-insulator transition. **(a)** Sheet resistance ($R_{sh}$) of film versus bias applied over $SiO_2$ gate dielectric at T = 4.2 K. $R_{sh}$ increases with decreasing $V_g$ / increasing $E_F$, indicating initial hole doping. The $R_{sh}$ value approaches a maximum at the Dirac point, which we posit to be near -150 V. Further electrostatic doping results in electrical breakdown. **(b)** $R_{sh}$ versus T at three $V_g$ levels indicates a transition from metallic behavior, where R increases with T, to insulating behavior, where R decreases with T. At this transition voltage, approximately -40 V, the Fermi level crosses the bulk valence band edge. **(c)** Schematic illustrating $E_F$ crossing the bulk valence band (BVB) edge at the metal-insulator transition voltage, overlaid on ARPES results for a similar $(Bi_{1-x}Sb_x)_2Te_3$ film, including the main features of the band structure. Inside the bulk gap are the two spin-polarized Dirac bands.

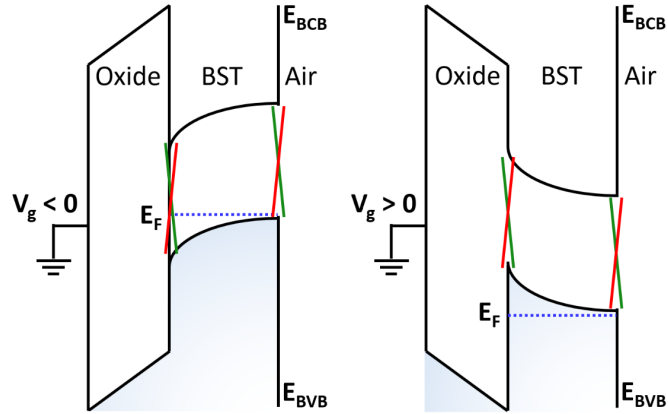

**Figure 4**: Schematic of energy bands across sample structure. Energy bands are shown for the silicon oxide, BST surfaces (green and red, at interfaces), BST bulk, and air. For negative bias, additional holes are accumulated in the BST layer. For positive bias, holes are depleted. The screening length is of order the width of the BST layer. As a result, the two topological surfaces likely experience different doping. Due to the much lower doping in the a-Te, band bending at the BST / a-Te interface occurs mostly in the a-Te and so is not shown here.

**Supplementary Information for: "Gate-Variable Mid-Infrared Optical Transitions in a $(Bi_{1-x}Sb_x)_2Te_3$ Topological Insulator"**

[1]William S. Whitney, [2,3]Victor W. Brar, [4]Yunbo Ou, [5,6]Yinming Shao, [2]Artur R. Davoyan, [5,6]D. N. Basov, [7]Ke He, [7]Qi-Kun Xue and [2]Harry A. Atwater

[1]Department of Physics, California Institute of Technology, Pasadena, California 91125, USA
[2]Thomas J. Watson Laboratory of Applied Physics, California Institute of Technology, Pasadena, California 91125, USA
[3]Kavli Nanoscience Institute, California Institute of Technology, Pasadena, California 91125, USA
[4]Beijing National Laboratory for Condensed Matter Physics, Institute of Physics, The Chinese Academy of Sciences, Beijing 100190, China
[5]Department of Physics, University of California-San Diego, La Jolla, California 92093, USA
[6]Department of Physics, Columbia University, New York, NY
[7]State Key Laboratory of Low-Dimensional Quantum Physics, Department of Physics, Tsinghua University, Beijing 100084, China

*Corresponding author: Harry A. Atwater (haa@caltech.edu)

**Epitaxial lift-off methodology:**

After spin-coating PMMA (950 A8) onto the surface of the films and baking them on a hot-plate at 170 C for 2 minutes, the chips are placed into a bath of buffered hydrofluoric acid. The film begins peeling off the substrate after 2-3 hours, at which point the chip is placed into a series of DI water baths. The chip is held at the surface of the water, and surface tension is used to complete peeling of the film. The film floats on the surface of the water, and is lifted out with a thermal oxide on silicon chip. This chip is dried overnight, and the PMMA is removed with acetone. This process and a transferred film are shown in Supplementary Fig. 1.

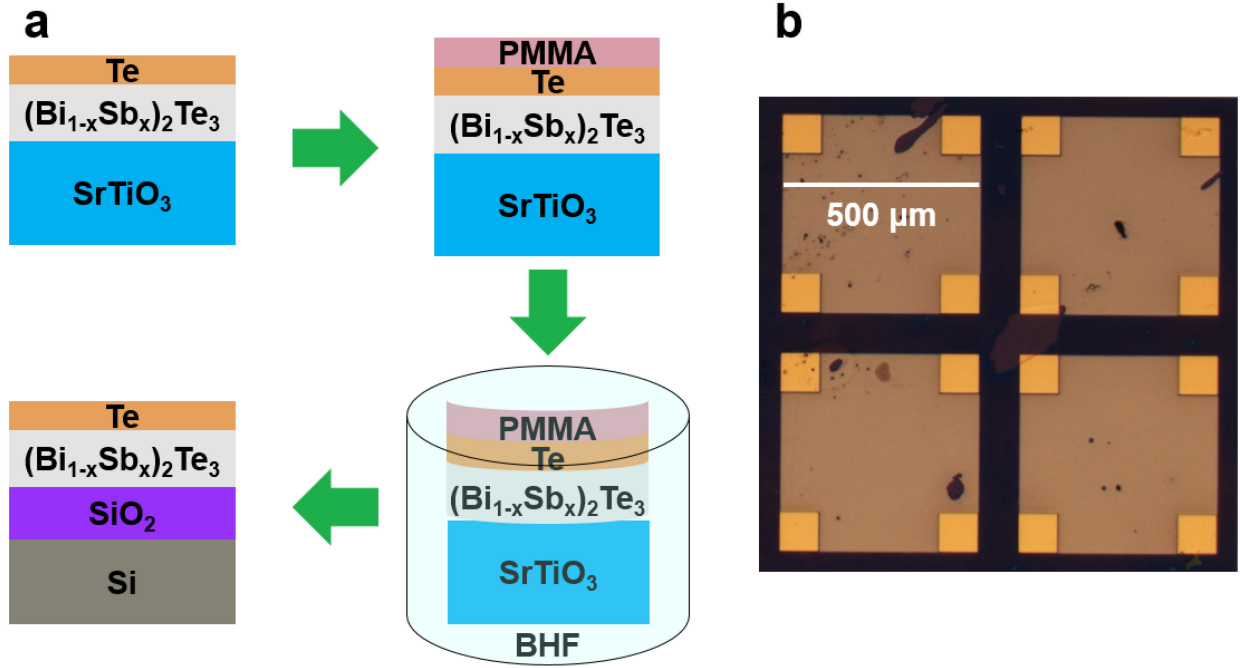

**Supplementary Figure 1**: Transfer method and result. **(a)** Outline of transfer process. PMMA is spin-coated onto $(Bi_{1-x}Sb_x)_2Te_3$ film on its STO growth substrate. The sample is then submerged into buffered HF until the PMMA / Te / $(Bi_{1-x}Sb_x)_2Te_3$ stack peels from the STO. The stack is scooped out of water with a $SiO_2$ / Si chip, dried, and treated with acetone to remove the PMMA. **(a)** Optical microscope image of devices fabricated on a film transferred to $SiO_2$ / Si.

**Low temperature transmittance:**

Gate-variable FTIR transmittance is also measured at 5 K, in order to understand the low temperature infrared response of the $(Bi_{1-x}Sb_x)_2Te_3$ film and look for TSS to TSS interband transitions. These measurements are performed in a modified Oxford cryostat using a Bruker Lumos infrared microscope and spectrometer, in the Basov Lab facilities at UCSD. The behavior seen – shown in Supplementary Fig. 2 – is consistent with that seen at 78 K, with a smaller Fermi-Dirac distribution width. An additional kink feature is seen near six microns, but no clear evidence of TSS to TSS interband transitions – which should show a universal optical conductivity of $\pi e^2/8h$ above $2E_F$ – is seen.[46] The persistence of bulk-related modulation at high gate voltage and low temperature indicates that some band bending is likely present. The presence of accumulating ice dampens the modulation around a narrow feature at three microns.[47]

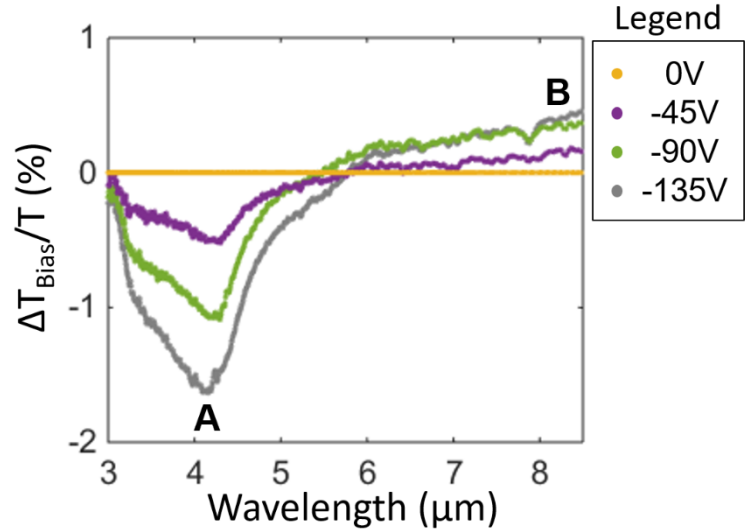

**Supplementary Figure 2**: Gate-variable FTIR transmittance at T = 5 K. Transmittance is shown normalized to zero bias case, as the Fermi level is pushed into the bulk band gap and towards the Dirac point. Similar behavior is seen as in T = 78 K, with two main differences. The band-edge Pauli-blocking effect is sharper and more pronounced, likely due to narrowing of the Fermi-Dirac distribution of carrier energies. There also appears a kink feature in transmittance near 6 microns. This feature may be related, but no clear evidence of TSS to TSS interband transitions is observed.

**Room temperature transmittance and reflectance:**

Gate-variable FTIR transmittance and reflectance are also measured at 300 K, in order to understand how the infrared response of the $(Bi_{1-x}Sb_x)_2Te_3$ film varies with temperature and demonstrate the possibility of room temperature optical modulation. These measurements are performed in the same configuration as the 78 K measurements. The behavior seen, shown in Supplementary Fig. 3 – is consistent with that seen in the 78 K measurements, but show less modulation.

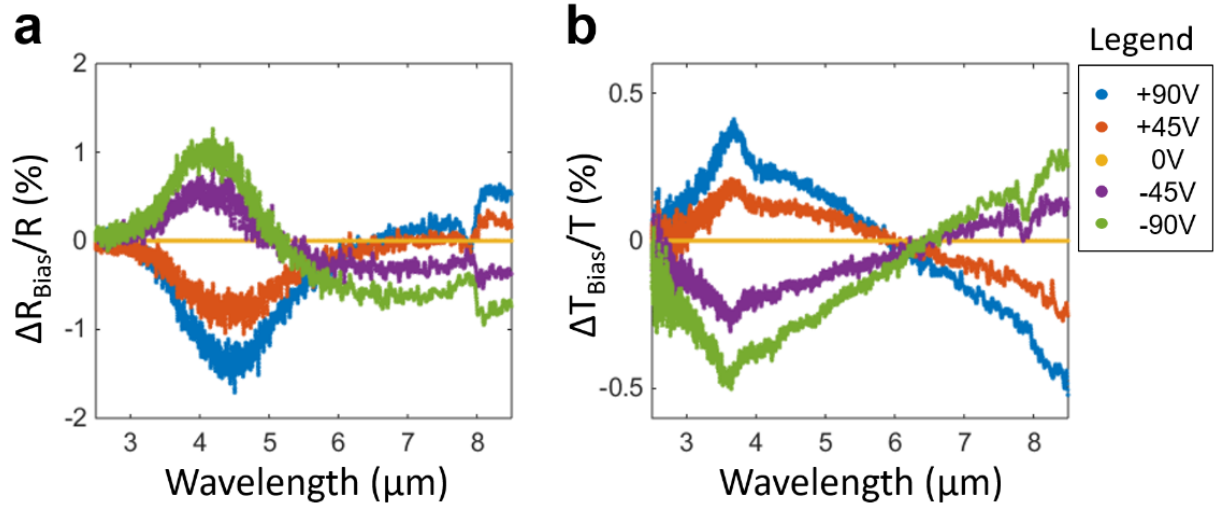

**Supplementary Figure 3:** Room temperature FTIR spectra. **(a)** Gate-variable FTIR transmittance, shown normalized to the zero bias case. **(b)** Gate-variable FTIR reflectance, shown normalized to the zero bias case. Similar behavior is seen as with 78 K measurements, but with a smaller modulation depth.